\documentclass{llncs}
\usepackage[english]{babel}
\usepackage{url}
\usepackage{graphicx}
\usepackage{amssymb}
\usepackage{fancyvrb}

\newcommand{\boxtt}[1]{\mbox{\small\texttt{#1}}}

\newif\iffinal 

\usepackage{tikz}
\pgfrealjobname{aqfmsiat} 
\iffinal
  
\else
  
\fi

\begin{document}
\hyphenation{rais--ing}
\hyphenation{SCSCP}


\pagestyle{empty}


\title{Integrating multiple sources to answer questions in Algebraic Topology.\thanks{Partially supported by Ministerio de Ciencia e Innovaci\'on, project MTM2009-13842-C02-01. The final publication of this paper is available at www.springerlink.com}}

\titlerunning{Answering questions from multiple sources, in Algebraic Topology}

\author{J\'onathan Heras \and Vico Pascual \and Ana Romero \and Julio Rubio}

\authorrunning{J. Heras et al.}

\institute{Departamento de Matem\'{a}ticas y Computaci\'{o}n,
Universidad de La Rioja,
\\ Edificio Vives, Luis de Ulloa s/n,
E-26004 Logro\~no (La Rioja, Spain).
\\ \email{\{jonathan.heras, vico.pascual, ana.romero, julio.rubio\}@unirioja.es}}

\maketitle

\sloppy

\begin{abstract}

We present in this paper an evolution of a tool from a user interface for a concrete Computer Algebra system for Algebraic Topology (the Kenzo system), to a front-end allowing the interoperability among different sources for computation and deduction. The architecture allows the system not only to interface several systems, but also to make them cooperate in shared calculations.

\end{abstract}

\section{Different questions, different sources}

When working in Mathematics, usually the researcher uses different sources of information. Typically, he can consult some papers or textbooks, make some computations with a Computer Algebra system, check the results against some known tables or, more rarely, verify some conjectures with a proof assistant tool. That is to say, Mathematical Knowledge is dispersed among several sources.

Our aim in this work is to mechanize, in some particular cases, the management of these multiple-source information systems. Since it would be too pretentious to try to solve fully this problem, we work in a very specific context. Thematically, we restrict ourselves to (a subset of) Algebraic Topology. With respect to the sources, in order to have a representation wide enough, we have chosen two Computer Algebra systems (Kenzo and GAP), a theorem prover (ACL2) and a small expert system developed by us. The objective of the expert system is computing \emph{homotopy} groups. Kenzo and GAP can compute \emph{homology} groups of different spaces, but the calculation of \emph{homotopy} is in general much harder. Our homotopy expert system tries to take profit of theoretical knowledge contained in theorems (tables have been excluded up to now, since they are considered less difficult to integrate, from a technological point of view), and can ask computational results to Kenzo, if needed.

This paper is a natural continuation of~\cite{HPR08} and~\cite{HPR09}. There are three main contributions in the paper: an architecture based on the Broker pattern~\cite{BHS07} (proven as an open, flexible and adaptable tool); an Homotopy Expert System (HES) that allows non-trivial computations (and explanations) interacting with Kenzo; and the automation of the interoperability between Kenzo and GAP.

From the symbolic computation literature, we looked for inspiration in different projects and frameworks such as the MathWeb software bus~\cite{MATHWEB-SB}, its successor the MathServe Framework~\cite{MATHWEB-f}, the MoNET project~\cite{cohen00,monet} or the MathBroker~\cite{MathBroker} and MathBroker II~\cite{MathBrokerII} projects, as well as in other works as~\cite{Fre08} or~\cite{Smi04}.

\section{General view of the system}

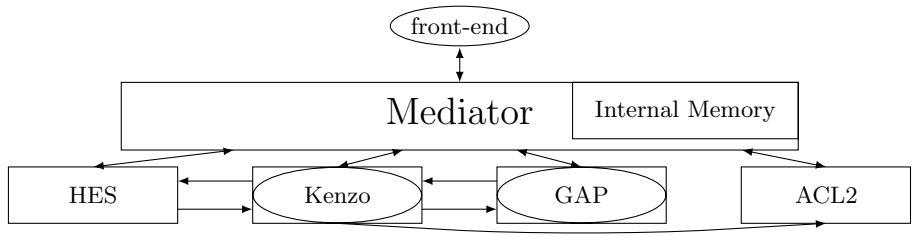
\begin{figure}
    \centering
    \beginpgfgraphicnamed{architecture-external}%
    \begin{tikzpicture}[scale=0.75]
\draw
(8,2.8) rectangle (20,4) 
(14,3.5) node {{\Large Mediator}} 
(16,3) rectangle (20,4) 
(18,3.5) node {Internal Memory} 
(14,5) ellipse (35pt and 10pt) 
(14,5) node {front-end} 
(6,1.5) rectangle (9,2.5) 
(7.5,2) node {{\small HES}} 
(10.33,1.5) rectangle (13.33,2.5) 
(11.83,2) ellipse (42.5pt and 14pt) 
(11.83,2) node {{\small Kenzo}} 
(14.66,1.5) rectangle (17.66,2.5) 
(16.16,2) ellipse (42.5pt and 14pt) 
(16.16,2) node {{\small GAP}} 
(18.99,1.5) rectangle (21.99,2.5) 
(20.5,2) node {{\small ACL2}} 
;
\draw [latex-latex] (14,4) -- (14,4.62); 
\draw [-latex] (9,1.75) -- (10.33,1.75); 
\draw [latex-] (9,2.25) -- (10.33,2.25); 
\draw [-latex] (13.33,1.75) -- (14.66,1.75); 
\draw [latex-] (13.33,2.25) -- (14.66,2.25); 
\draw [rounded corners=50pt,-latex] (11.83,1.5) -- (16.16,1.25) -- (20.5,1.5); 
\draw [latex-latex] (7.5,2.5) -- (10,2.8); 
\draw [latex-latex] (11.83,2.5) -- (13,2.8); 
\draw [latex-latex] (16.16,2.5) -- (15,2.8); 
\draw [latex-latex] (20.5,2.5) -- (19,2.8); 
\end{tikzpicture}%
    \endpgfgraphicnamed%

\vskip -10pt
\caption{Broker architectural pattern}\label{broker}
\end{figure}

The \emph{Broker} architectural pattern~\cite{BHS07} can be used to structure software systems with decoupled components that interact through service invocations. The Broker pattern defines three kinds of participants: \emph{clients}, \emph{components}, and the \emph{broker} itself. A scheme of our architecture based on this pattern is depicted in Figure~\ref{broker}. The mediator (broker) component embeds an \emph{Internal Memory} where a strategy of \emph{memoization} has been systematically implemented (based on the same idea used in GAP for attributes, see~\cite{GAP-tutorial}). The system stores the results in the internal memory when a computation is executed for the first time, and if the same computation is asked again later, the result is simply looked up and returned, without further computation.

The decorator pattern~\cite{BHS07} is used to wrap objects of our system with information, like the type of the object (simplicial set, group,\ldots) or the reduction degree~\cite{HPR08} if the object is a topological space. This information guides the mediator to decide which component to use. Namely, Kenzo~\cite{Kenzo} (a Symbolic Computation system devoted to Algebraic Topology) is the core for computations related to homology groups of spaces, GAP~\cite{GAP} (a Computer Algebra system in the area of Computational Group Theory) and HAP~\cite{HAP} (a GAP homological algebra library) are the core for computations related to group homology, ACL2~\cite{ACL2} (a first order logic theorem prover) is the kernel for verifying the correctness of statements and, finally, the Homotopy Expert System (a small module developed by us described in the next paragraph and from now on called HES) is in charge of computing homotopy groups.

HES is a rule-based expert system. The structure of a rule-based expert system, see~\cite{GR05}, consists of, and the HES is no exception, the following components: the \emph{Working memory (the facts)}, the \emph{Knowledge base (the rules)}, the \emph{Inference engine},  a \emph{Knowledge acquisition} module and an \emph{Explanation facility} module. In the scope of the HES, the facts are properties associated with the objects (for instance, ``$\forall n\in \mathbb{N}:~~ \Delta^n \textrm{ is a contractible space}$''). The current knowledge base is made up of 23 rules (such as, ``\boxtt{if} $X$ is contractible \boxtt{and} $n\geq 1$ \boxtt{then} $\pi_n(X)=0$''). The inference engine uses the \emph{forward chaining} method for reasoning, see~\cite{GR05}. To grapple with the knowledge acquisition aspects, the HES takes profit from both the RuleML markup language~\cite{RuleML} and the OMDoc format~\cite{OMDoc}, the former one is used to specify rules in a declarative way and the second one to store concrete functionalities. Last but not least, gathering the applied rules and the facts that decorate each object, our HES is able to provide a trace containing the reasoning followed by it in order to reach a conclusion.

However, the power of our system does not lie in gathering several computer algebra systems and theorem provers and use them separately with the same front end, but interconnecting them to reach new results. The communication among modules is performed by means of the OpenMath language~\cite{OpenMath}, used to represent the objects in a common language for all the systems.

In~\cite{RER09} an approach to coordinate GAP and Kenzo was presented. In that work GAP and Kenzo cooperate in order to compute homology groups of some spaces. These spaces with their homology can then be used in other constructions and applications. Some enhancements of that tool provided by our system are: avoidance of the installation of several programs and packages, automation of communication steps (here the SCSCP protocol~\cite{SCSCP} plays a key role) and concealment of the details to mix the systems.

The general procedure and technology to connect with the ACL2 system explained in~\cite{HPR09} is now applied to the context of group homology. The Common Lisp code used in Kenzo to represent a group is sent to ACL2 as an instance of an ACL2 \emph{encapsulate} (a mechanism to introduce new functions symbols by axioms constraining them to have certain properties) by means of our broker, which is also in charge of invoking ACL2 in a way transparent for the user.

In another line, Kenzo and the HES cooperate to compute homotopy groups of spaces. In this case, the HES requests Kenzo to compute homology groups which can be used to obtain homotopy groups. Whereas Kenzo communicates with the HES in order to send it results. The idea consists of gathering the knowledge stored in the HES and chaining several tools available in Kenzo to get results which are not reachable by anyone of them working in an independent way.

\section{Putting all together}\label{pat}

\begin{figure}
    \centering
    \includegraphics[width=0.5\columnwidth]{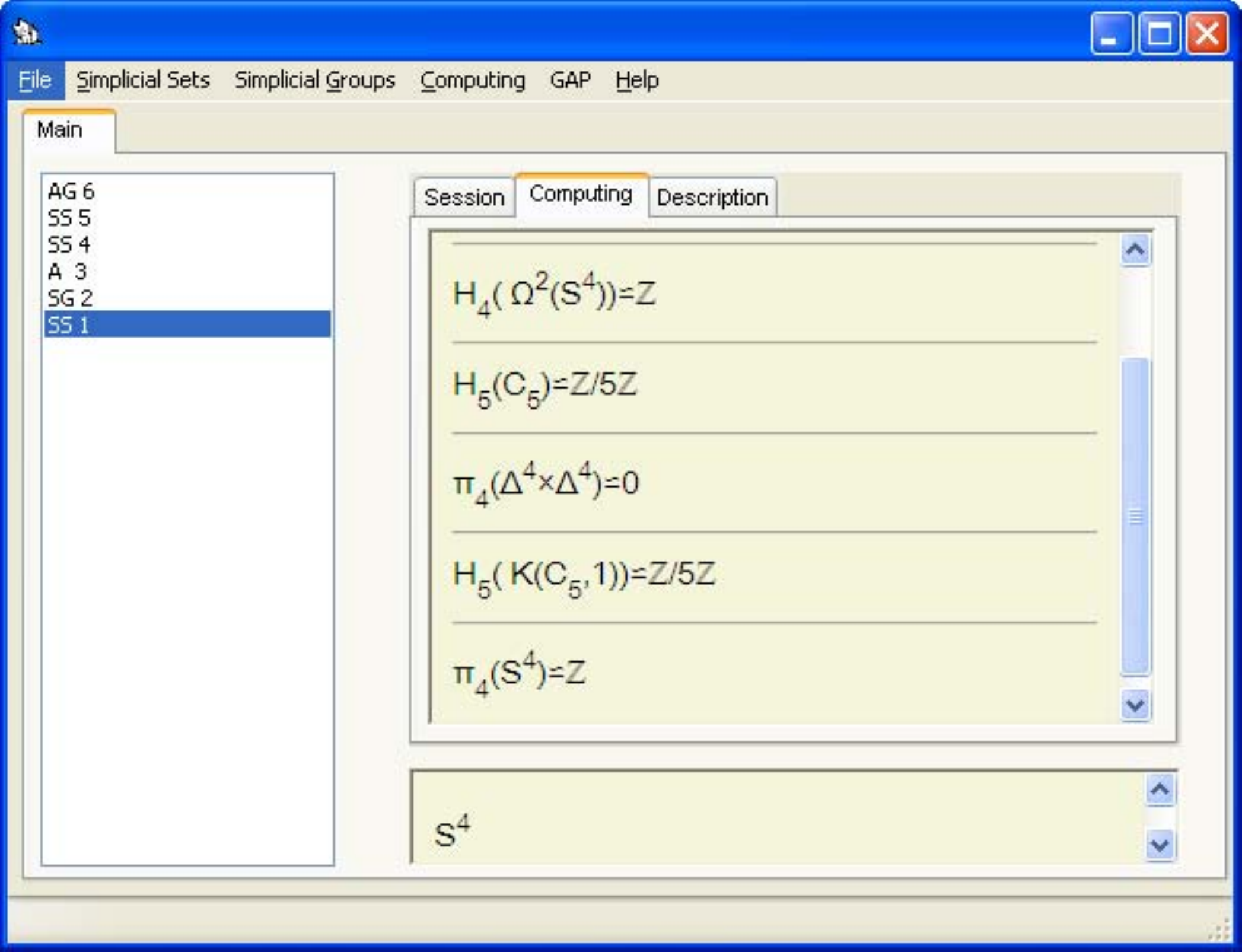}
\vskip -10pt
\caption{The front end for computations}\label{screenshot}
\end{figure}

Our current front-end has evolved from the user interface for Kenzo presented in~\cite{HPR08}. Its presentation layer is kept, but its internal mediator has been enriched to support different sources of information. Figure~\ref{screenshot} displays some computations which took profit of the following interactions: $H_4(\Omega^2(S^4))=\mathbb{Z}$ (computed with Kenzo), $H_5(C_5)=\mathbb{Z}/5\mathbb{Z}$ (computed with GAP), $\pi_4(\Delta^4\times \Delta^5)=0$ (obtained with the Homotopy Expert System), $H_5(K(C_5,1))=\mathbb{Z}/5\mathbb{Z}$ (computed with Kenzo + GAP), $\pi_4(S^4)=\mathbb{Z}$ (obtained with the Homotopy Expert System + Kenzo).

It is worth noting that all results are shown to the user in a unique screen and that the computations are asked from a sole menu, then the user does not know the system in charge of computing neither the collaboration among computer algebra systems, he only receives the desired result. The technical details are hidden to the user. The results related to ACL2 are shown in a different tab to split the computations from the deductions.

\section{Conclusions and Further Work}

In this paper an architecture to integrate different tools for computing and logical reasoning in Algebraic Topology is presented. Even if our proposal has a limited extend, both thematically and from the point of view of the core systems, we think it shows a solid line of research that could be exported to other areas of mathematical knowledge management. OpenMath technologies are the essential tool ensuring the interoperability among systems (even integration in some cases). This interoperability has a vertical dimension (from the mediator to the kernel systems) as well as a horizontal axis (allowing direct interconnection of kernel systems). The modules can be taken from their sources (as in the cases of Kenzo and ACL2), invoked in a remote manner (like the GAP server, connected via the SCSCP protocol) or even developed in an ad-hoc way (as our Homotopy Expert System).

Several research lines are still open. The most important ones are related to giving more resources to the user to manage the interaction. Moreover, it would be also necessary to improve the interaction with the ACL2 system. At this moment the queries must be pre-processed; a comfortable way of introducing questions about the truth of properties of intermediary objects, dynamically generated during a computing session, should be provided. Last, and the most difficult one, a meta-language should be designed to specify how and when a new kernel system can be plugged in the framework. This capability and the necessity of orchestrating the different services suppose a real challenge, which will be explored by means of OpenMath technologies.

\bibliographystyle{plain}
\bibliography{aqfmsiat}

\end{document}